\documentclass[conference]{IEEEtran}
\ifCLASSINFOpdf
  \usepackage[pdftex]{graphicx}
\else
\fi
\usepackage{amsmath,amsfonts}

\begin{document}
%
\title{
Mathematical Foundations of the GraphBLAS}

\author{\IEEEauthorblockN{
  Jeremy Kepner (MIT Lincoln Laboratory Supercomputing Center),
  Peter Aaltonen (Indiana University),\\
  David Bader (Georgia Institute of Technology),
  Ayd\i{}n Bulu{\c{c}} (Lawrence Berkeley National Laboratory),\\
  Franz Franchetti (Carnegie Mellon University),
  John Gilbert (University of California, Santa Barbara),\\
  Dylan Hutchison (University of Washington),
  Manoj Kumar (IBM),\\
  Andrew Lumsdaine (Indiana University),
  Henning Meyerhenke (Karlsruhe Institute of Technology),\\
  Scott McMillan (CMU Software Engineering Institute),
  Jose Moreira (IBM),\\
  John D. Owens (University of California, Davis),
  Carl Yang (University of California, Davis),\\
  Marcin Zalewski (Indiana University),
  Timothy Mattson (Intel)
}
}


%


\maketitle

\begin{abstract}
The GraphBLAS standard ({\sf GraphBlas.org}) is being developed to bring the potential of matrix-based graph algorithms to the broadest possible audience.  Mathematically, the GraphBLAS defines a core set of matrix-based graph operations that can be used to implement a wide class of graph algorithms in a wide range of programming environments.  This paper provides an introduction to the mathematics of the GraphBLAS.  Graphs represent connections between vertices with edges.  Matrices can represent a wide range of graphs using adjacency matrices or incidence matrices.  Adjacency matrices are often easier to analyze while incidence matrices are often better for representing data.  Fortunately, the two are easily connected by matrix multiplication. A key feature of matrix mathematics is that a very small number of matrix operations can be used to manipulate a very wide range of graphs.  This composability of a small number of operations is the foundation of the GraphBLAS.  A standard such as the GraphBLAS can only be effective if it has low performance overhead.  Performance measurements of prototype GraphBLAS implementations indicate that the overhead is low.
\end{abstract}


%
\IEEEpeerreviewmaketitle

\section{Introduction}
\let\thefootnote\relax\footnotetext{This material is based in part upon work supported by the NSF under grant number DMS-1312831, by DOE ASCR under contract number DE-AC02-05CH11231, and by the DoD under contract number FA8721-05-C-0003.  Any opinions, findings, and conclusions or recommendations expressed in this material are those of the authors and do not necessarily reflect the views of the National Science Foundation, the Department of Energy, or the Department of Defense.}  Graphs are among the most important abstract data structures in computer science, and the algorithms that operate on them are critical to applications in bioinformatics~\cite{Georganas2014}, computer networks, and social media \cite{Ediger2010,Ediger2011,Riedy2012,RiedyBader2013}.  Graphs have been shown to be powerful tools for modeling complex problems because of their simplicity and generality \cite{Staudt2016,BergaminiMeyerhenke2016}.  For this reason, the field of graph algorithms has become one of the pillars of theoretical computer science, informing research in such diverse areas as combinatorial optimization, complexity theory, and topology.  Graph algorithms have been adapted and implemented by the military, commercial industry, and researchers in academia, and have become essential in controlling the power grid, telephone systems, and, of course, computer networks.

Parallel graph algorithms are notoriously difficult to implement and optimize \cite{Ediger2012,EdigerBader2013,McLaughlinBader2014a,McLaughlinBader2014b,McLaughlin2014,StaudtMeyerhenke2016}.  The irregular data access patterns and inherently high communication-to-computation ratios found in graph algorithms mean that even the best algorithms will have parallel efficiencies that decrease as the number of processors is increased \cite{BulucGilbert2012,Azad2015}. Recent work on communication-avoiding algorithms, and their applications to graph computations \cite{Ballard2013,Solomonik2013}, might defer but not completely eliminate the parallel scalability bottleneck. Consequently, novel hardware architectures will also be required \cite{Song2010,Song2013}.  A common graph processing interface provides a useful tool for optimizing both software and hardware to provide high performance graph applications.

  The duality between the canonical representation of graphs as abstract collections of vertices and edges and a matrix representation has been a part of graph theory since its inception \cite{Konig1931, Konig1936}.  Matrix algebra has been recognized as a useful tool in graph theory for nearly as long (see \cite{Harary1969} and references therein, in particular \cite{Sabadusi1960,Weischel1962,McAndrew1963, TehYap1964,McAndrew1965,HararyTauth1966,Brualdi1967}).  The modern description of the duality between graph algorithms and matrix mathematics (or sparse linear algebra) has been extensively covered in the literature and is summarized in the cited text \cite{KepnerGilbert2011}.  This text has further spawned the development of the GraphBLAS math library standard (GraphBLAS.org)\cite{Mattson2013} that has been developed in a series of proceedings \cite{Mattson2014a,Mattson2014b,Mattson2015,Buluc2015,Mattson2016} and implementations \cite{BulucGilbert2011,Kepner2012,Ekanadham2014,Hutchison2015,Anderson2016,Zhang2016}.  This paper describes the mathematical properties that have been developed since \cite{KepnerGilbert2011} to support the GraphBLAS. 

  The foundational mathematical concepts for matrix-based graph analysis are the adjacency matrix and incidence matrix representations of graphs.  From these concepts, a more formal definition of a matrix can be constructed.  How such a matrix can be manipulated depends on the types of values the matrix holds and the operations allowed on those values.  Furthermore, the mathematical properties of the matrix values determine the mathematical properties of the whole matrix.  This paper describes the key mathematical concepts of the GraphBLAS and presents preliminary results that show the overhead of the GraphBLAS is minimal (as compared to their underlying matrix libraries).

\section{Adjacency Matrix}
\begin{figure}[htb]
  \centering
    \includegraphics[width=3in]{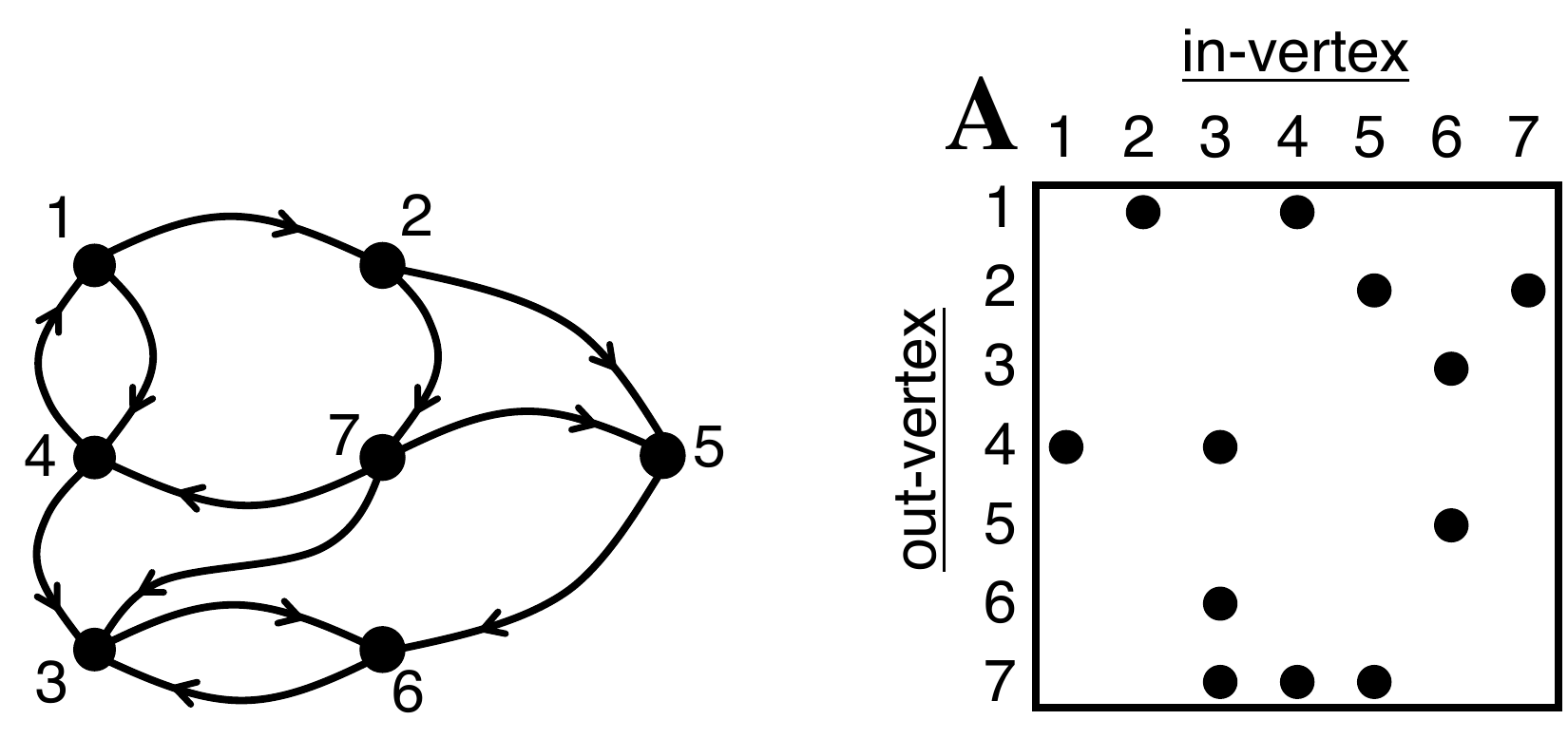}
      \caption{(left) Seven-vertex graph with 12 edges.  Each vertex is labeled with an integer.  (right)  $7 \times 7$ adjacency matrix $\mathbf{A}$ representation of the graph. $\mathbf{A}$ has 12 nonzero entries corresponding to the edges in the graph.}
      \label{fig:AdjacencyMatrix}
\end{figure}

Given an adjacency matrix $\mathbf{A}$, if
$$
  \mathbf{A}(i,j) = 1
$$
then there exists an edge going from vertex $i$ to vertex $j$ (see Figure~\ref{fig:AdjacencyMatrix}).  Likewise, if
$$
  \mathbf{A}(i,j) = 0
$$
then there is no edge from $i$ to $j$.  Adjacency matrices can have direction, which means that $\mathbf{A}(i,j)$ may not be the same as $\mathbf{A}(j,i)$.  Adjacency matrices can also have edge weights.  If
$$
  \mathbf{A}(i,j) = a \neq 0
$$
then the edge going from $i$ to $j$ is said to have weight $a$. Adjacency matrices provide a simple way to represent the connections between vertices in a graph.  Adjacency matrices are often square, and both the out-vertices (rows) and the in-vertices (columns) are the same set of vertices.  Adjacency matrices can be rectangular, in which case the out-vertices (rows) and the in-vertices (columns) are different sets of vertices.  Such graphs are often called bipartite graphs.  In summary, adjacency matrices can represent a wide range of graphs, which include any graph with any set of the following properties: directed, weighted, and/or bipartite.

\section{Incidence Matrix}
\begin{figure}[!htb]
  \centering
    \includegraphics[width=3in]{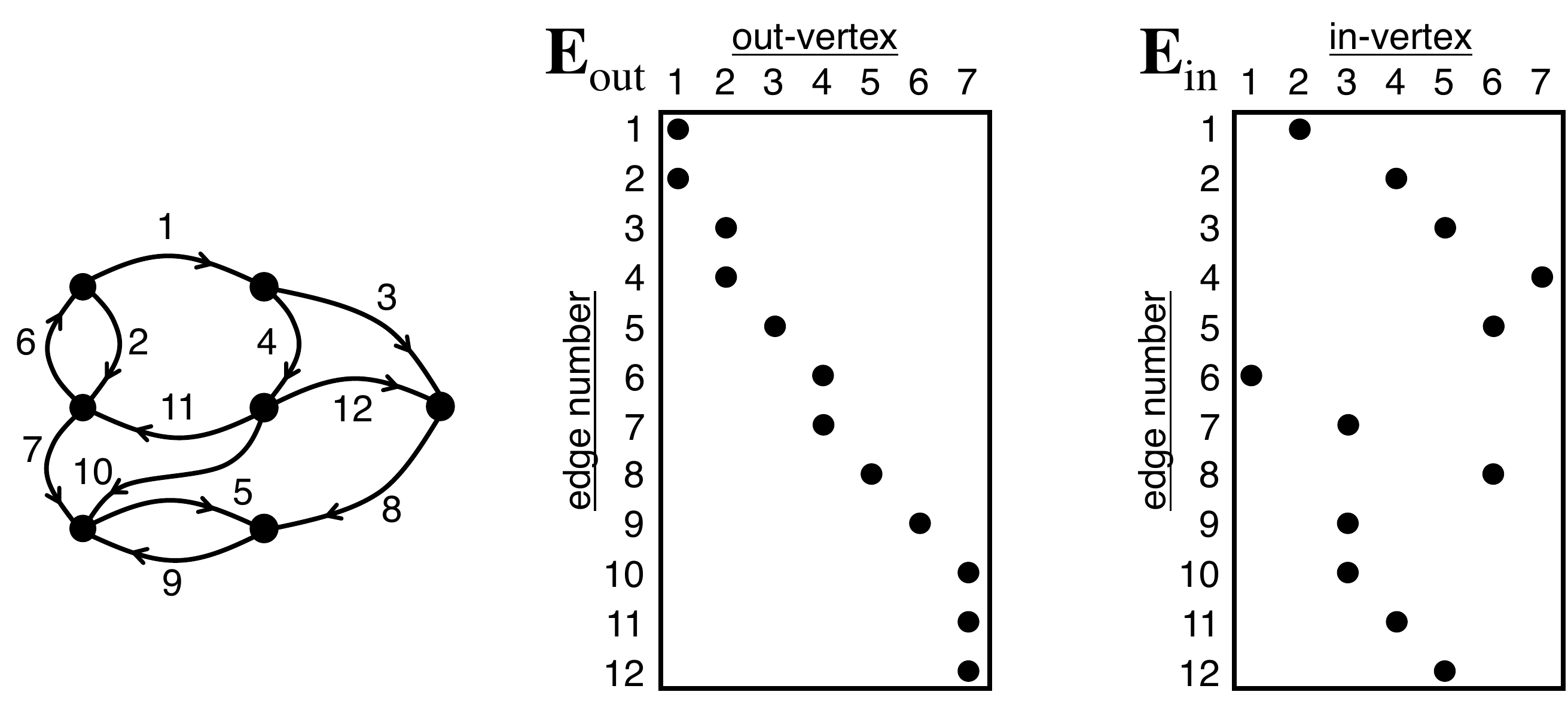}
      \caption{(left) Seven-vertex graph with 12 edges.  Each edge is labeled with an integer; the vertex labels are the same as in Figure~\ref{fig:AdjacencyMatrix}.  (middle)  $12 \times 7$ incidence matrix $\mathbf{E}_{\rm out}$ representing the out-vertices of the graph edges.   (right)  $12 \times 7$ incidence matrix $\mathbf{E}_{\rm in}$ representing the in-vertices of the graph edges. Both $\mathbf{E}_{s\rm tart}$ and $\mathbf{E}_{\rm in}$ have 12 nonzero entries corresponding to the edges in the graph.}
      \label{fig:IncidenceMatrix}
\end{figure}

\begin{figure}[!htb]
  \centering
    \includegraphics[width=3in]{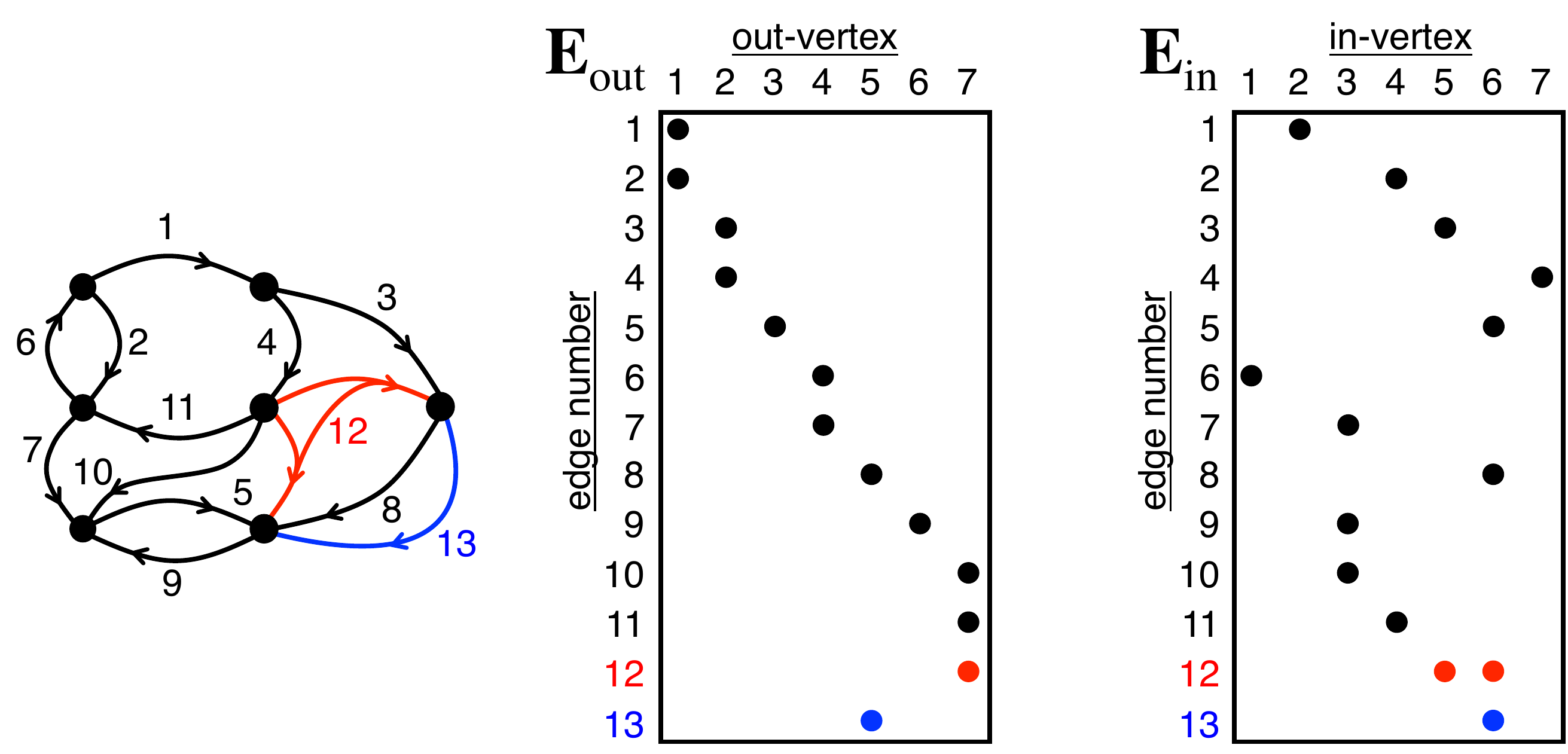}
      \caption{Graph and incidence matrices from Figure~\ref{fig:IncidenceMatrix} with a hyper-edge (edge 12) and a multi-edge (edge 13).  The graph is a hyper-graph because edge 12 has more than one in-vertex.  The graph is a multi-graph because edge 8 and edge 13 have the same out- and in-vertex.}
      \label{fig:IncidenceMatrixMultiHyper}
\end{figure}

 An incidence, or edge matrix $\mathbf{E}$, uses the rows to represent every edge in the graph and the columns to represent every vertex.  There are a number of conventions for denoting an edge in an incidence matrix.  One such convention is to use two incidence matrices
$$
  \mathbf{E}_{\rm out}(k,i) = 1 ~~~~~ {\rm and} ~~~~~~ \mathbf{E}_{\rm in}(k,j) = 1
$$
to indicate that edge $k$ is a connection from $i$ to $j$ (see Figure~\ref{fig:IncidenceMatrix}).  Incidence matrices are useful because they can easily represent multi-graphs, hyper-graphs, and multipartite graphs.  These complex graphs are difficult to capture with an adjacency matrix.  A multi-graph has multiple edges between the same vertices.  If there was another edge, $k'$, from $i$ to $j$, this relationship can be captured in an incidence matrix by setting
$$
  \mathbf{E}_{\rm out}(k',i) = 1 ~~~~~ {\rm and} ~~~~~~ \mathbf{E}_{\rm in}(k',j) = 1
$$
(see Figure~\ref{fig:IncidenceMatrixMultiHyper}) [Note: Another convention is to use +1 and -1, in which case the resulting matrix multiplication is the graph Laplacian.]  In a hyper-graph, one edge can connect more than two vertices.  For example, to denote that edge $k$ has a connection from $i$ to $j$ and $j'$ can be accomplished by also setting
$$
  \mathbf{E}_{\rm in}(k,j') = 1
$$
(see Figure~\ref{fig:IncidenceMatrixMultiHyper}).  Furthermore, $i$, $j$, and $j'$ can be drawn from  different classes of vertices.  $\mathbf{E}$ can be used to represent multipartite graphs by defining an additional incidence array $\mathbf{E}'_{\rm in}$ and seting
$$
  \mathbf{E}'_{\rm in}(k,j') = 1
$$
Thus, an incidence matrix can be used to represent a graph with any set of the following graph properties: directed, weighted, multipartite, multi-edge, and/or hyper-edge.


\section{Matrix Values}
  A typical matrix has $m$ rows and $n$ columns of real numbers.  Such a matrix can be denoted as
$$
  \mathbf{A}: \mathbb{R}^{m \times n}
$$
The row and and column indexes of the matrix $\mathbf{A}$ are
$$
  i \in I = \{1,\ldots,m\}
$$
and
$$
  j \in J = \{1,\ldots,n\}
$$
so that any particular value $\mathbf{A}$ can be denoted as $\mathbf{A}(i,j)$.   The row and column indices of matrices are natural numbers $I,J : \mathbb{N}$.  [Note: a specific \emph{implementation} of these matrices might use IEEE 64-bit double-precision floating point numbers to represent real numbers, 64-bit unsigned integers to represent row and column indices, and the compressed sparse rows (CSR) format or the compressed sparse columns (CSC) format to store the nonzero values inside the sparse matrix.]

  A matrix of complex numbers
$$
  \mathbb{C} = \{x + y\sqrt{\text{-}1} : x,y \in \mathbb{R}\}
$$
is denoted
$$
  \mathbf{A}: \mathbb{C}^{m \times n}
$$
  A matrix of integers
$$
  \mathbb{Z} = \{\ldots, -1, 0, 1, \ldots\}
$$
is denoted
$$
  \mathbf{A}: \mathbb{Z}^{m \times n}
$$
  A matrix of natural numbers
$$
  \mathbb{N} = \{1, 2, 3, \ldots\}
$$
is denoted
$$
  \mathbf{A}: \mathbb{N}^{m \times n}
$$
Using the above concepts, a matrix is defined as the following two-dimensional (2D) mapping
$$
  \mathbf{A} : I \times J \rightarrow \mathbb{S}
$$
where the indices $I, J : \mathbb{Z}$ are finite sets of integers with $m$ and $n$ elements, respectively, and
$$
  \mathbb{S} \in \{\mathbb{R},\mathbb{Z},\mathbb{N}, \ldots \}
$$
is a set of scalars.  Without loss of generality, matrices can be denoted
$$
  \mathbf{A}: \mathbb{S}^{m \times n}
$$
A \emph{vector} is a matrix in which either $m=1$ or $n=1$. A column vector is denoted
$
 \mathbf{v} : \mathbb{S}^{m \times 1}
$
or simply
$
 \mathbf{v} : \mathbb{S}^{m \times 1}
$.
A row vector can be denoted
$
  \mathbf{v} : \mathbb{S}^{1 \times n}
$
or simply 
$
  \mathbf{v} : \mathbb{S}^{n}
$.
A scalar is a single element of a set
$
  s \in \mathbb{S}
$
and has no matrix dimensions.

\section{Scalar Operations}
Matrix operations are built on top of scalar operations that can be used for combining and scaling graph edge weights.  The primary scalar operations are standard arithmetic addition, such as
$$
  1 + 1 = 2
$$
and arithmetic multiplication, such as
$$
  2 \times 2 = 4
$$
These scalar operations of addition and multiplication can be defined to be a wide variety of functions.  To prevent confusion with standard arithmetic addition and arithmetic multiplication, $\oplus$ will be used to denote scalar addition and $\otimes$ will be used to denote scalar multiplication.  In this notation, standard arithmetic addition and arithmetic multiplication of real numbers
$$
  a, b, c \in \mathbb{R}
$$
where
$$
  \oplus \equiv + ~~~~~ \text{and} ~~~~~ \otimes \equiv \times
$$
results in
$$
   c = a \oplus b  ~~~~~~~~~ \Rightarrow ~~~~~~~~~ c = a + b
$$
and
$$
   c = a \otimes b  ~~~~~~~~~ \Rightarrow ~~~~~~~~~ c = a \times b
$$
Generalizing $\oplus$ and $\otimes$ to a variety of operations enables a wide range of algorithms on scalars of all different types (not just real or complex numbers).

Certain $\oplus$ and $\otimes$ combinations  over certain sets of scalars are particularly useful because they preserve essential mathematical properties, such as
additive commutativity
$$
  a \oplus b = b \oplus a
$$
multiplicative commutativity
$$
  a \otimes b = b \otimes a
$$
additive associativity
$$
  (a \oplus b) \oplus c = a \oplus (b \oplus c)
$$
multiplicative associativity
$$
  (a \otimes b) \otimes c = a \otimes (b \otimes c)
$$
and the distributivity of multiplication over addition
$$
  a \otimes (b \oplus c)  = (a \otimes b) \oplus (a \otimes c)
$$

The properties of commutativity, associativity, and distributivity are \emph{extremely} useful properties for building graph applications because they allow the builder to swap operations without changing the result.  Example combinations of $\oplus$ and $\otimes$ that preserve scalar commutativity, associativity, and distributivity include (but are not limited to) standard arithmetic
$$
  \oplus \equiv + ~~~~~~~~~ \otimes \equiv \times ~~~~~~~~~ a, b, c \in \mathbb{R}
$$
max-plus algebras
$$
  \oplus \equiv \max ~~~~~~~~~ \otimes \equiv + ~~~~~~~~~ a, b, c \in \{-\infty \cup \mathbb{R}\}
$$
max-min algebras
$$
  \oplus \equiv \max ~~~~~~~~~ \otimes \equiv \min ~~~~~~~~~ a, b, c \in \{\text{-}\infty \cup \mathbb{R}_{\leq 0} \}
$$
finite (Galois) fields such as GF(2)
$$
  \oplus \equiv {\rm xor} ~~~~~~~~~ \otimes \equiv {\rm and} ~~~~~~~~~ a, b, c \in \{0,1\}
$$
and power set algebras
$$
  \oplus \equiv \cup ~~~~~~~~~ \otimes \equiv \cap ~~~~~~~~~ a, b, c \subset \mathbb{Z}
$$
Other functions that do not preserve the above properties can also be defined for $\oplus$ and $\otimes$.  For example, it is often useful for $\oplus$ or $\otimes$ to pull in other data, such as vertex indices of a graph.

\section{Matrix Properties}
\label{MatrixProperties}
Associativity, distributivity, and commutativity are very powerful properties that enable the construction of composable graph algorithms (i.e., operations can be reordered with the knowledge that the answers will remain unchanged).  Composability makes it easy to build a wide range of graph algorithms with just a few functions.  Given matrices
$$
  \mathbf{A}, \mathbf{B}, \mathbf{C} \in \mathbb{S}^{m \times n}
$$
let their elements be specified by
$$
  a = \mathbf{A}(i,j) ~~~~~ 
  b = \mathbf{B}(i,j) ~~~~~
  c = \mathbf{C}(i,j)
$$
Commutativity, associativity, and distributivity of scalar operations translates into similar properties on matrix operations in the following manner.

\begin{figure}[htb]
  \centering
    \includegraphics[width=3in]{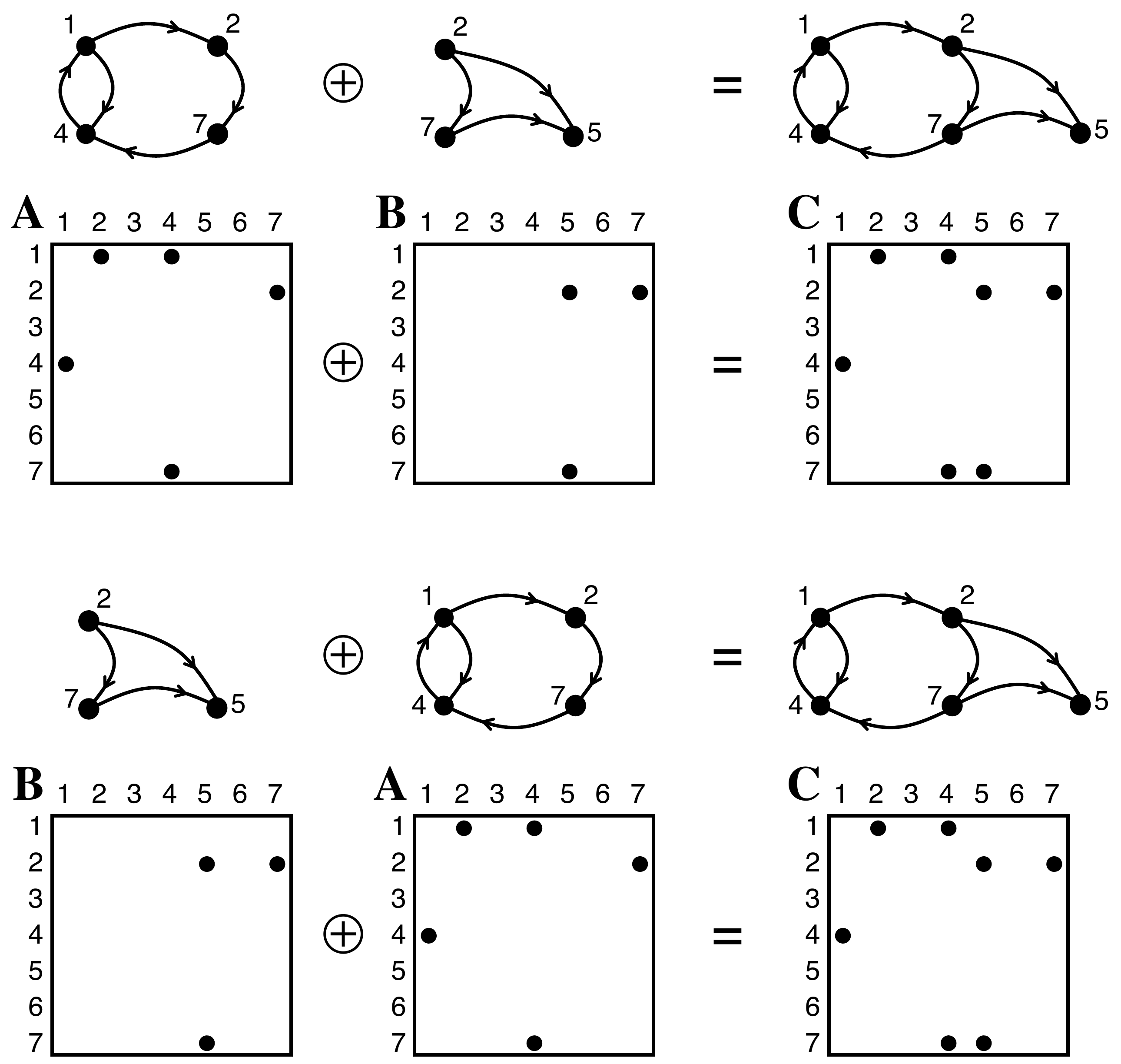}
      \caption{Illustration of the commutative property of the element-wise addition of two graphs and their corresponding adjacency matrix representations.}
      \label{fig:AdjacencyMatrixAdd}
\end{figure}

Additive commutativity allows graphs to be swapped and combined via matrix element-wise addition (see Figure~\ref{fig:AdjacencyMatrixAdd}) without changing the result
  $$
      a \oplus b = b \oplus a  ~~~~~ \Rightarrow ~~~~~
      \mathbf{A} \oplus \mathbf{B} = \mathbf{B} \oplus \mathbf{A}
  $$
  where matrix element-wise addition is given by
  $$
      \mathbf{C}(i,j) = \mathbf{A}(i,j) \oplus \mathbf{B}(i,j)
  $$
     
\begin{figure}[!htb]
  \centering
    \includegraphics[width=3in]{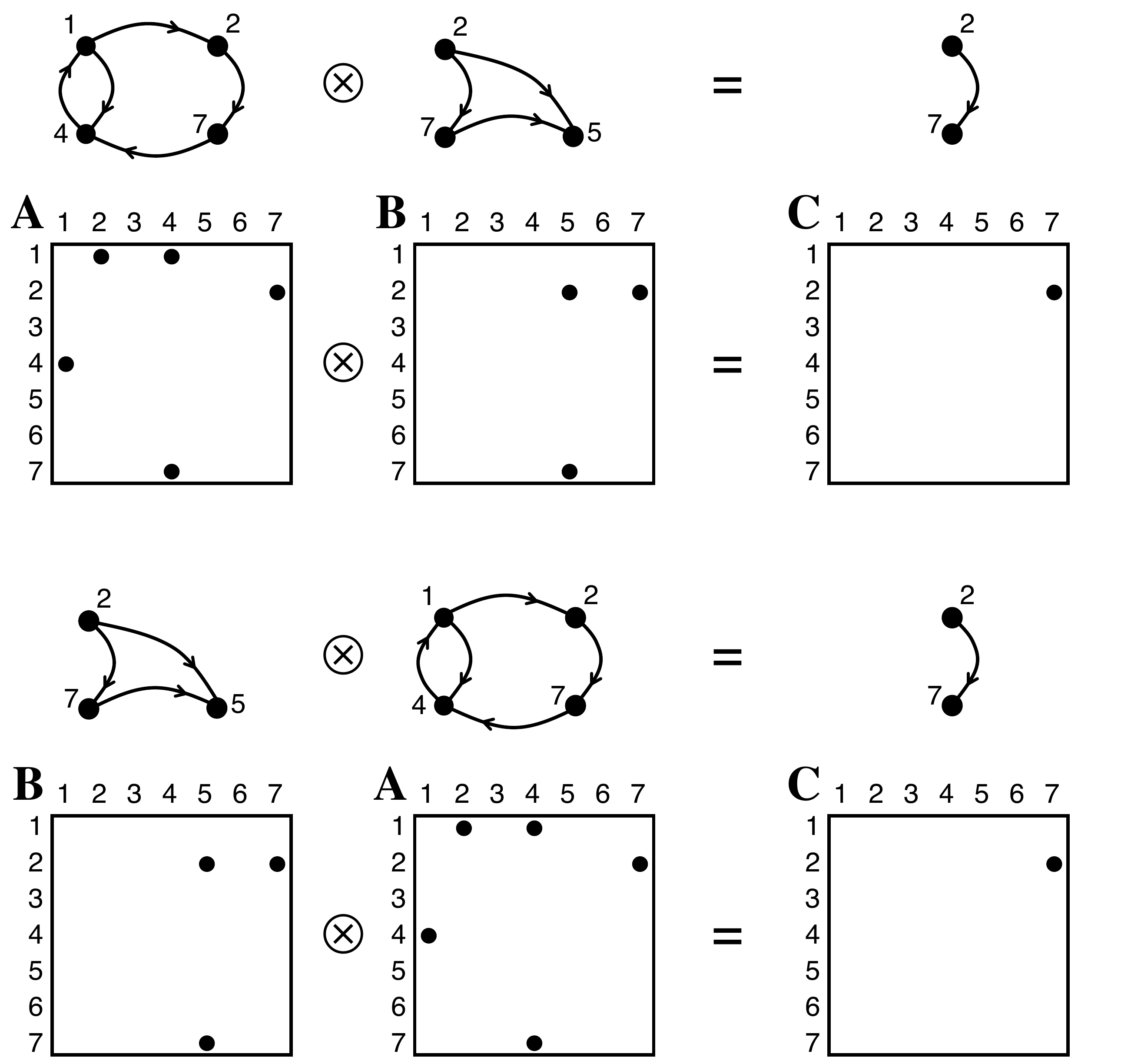}
      \caption{Illustration of the commutative property of the element-wise multiplication of two graphs and their corresponding adjacency matrix representations.}
      \label{fig:AdjacencyMatrixMult}
\end{figure}

Multiplicative commutativity allows graphs to be swapped, intersected, and scaled via matrix element-wise multiplication (see Figure~\ref{fig:AdjacencyMatrixMult}) without changing the result
  $$
      a \otimes b = b \otimes a  ~~~~~ \Rightarrow ~~~~~
      \mathbf{A} \otimes \mathbf{B} = \mathbf{B} \otimes \mathbf{A}
  $$
    where matrix element-wise (Hadamard) multiplication is given by
  $$
       \mathbf{C}(i,j) = \mathbf{A}(i,j) \otimes \mathbf{B}(i,j)
  $$

Additive associativity allows graphs to be combined via matrix element-wise addition in any grouping without changing the result
  $$
      (a \oplus b) \oplus c = a \oplus (b \oplus c)   ~~~ \Rightarrow ~~~
      (\mathbf{A} \oplus \mathbf{B}) \oplus \mathbf{C} = \mathbf{A} \oplus (\mathbf{B} \oplus \mathbf{C})
  $$

Multiplicative associativity allows graphs to be intersected and scaled via matrix element-wise multiplication in any grouping without changing the result
  $$
      (a \otimes b) \otimes c = a \otimes (b \otimes c)   ~~~ \Rightarrow ~~~
      (\mathbf{A} \otimes \mathbf{B}) \otimes \mathbf{C} = \mathbf{A} \otimes (\mathbf{B} \otimes \mathbf{C})
  $$

Element-wise distributivity allows graphs to be intersected and/or scaled and then combined or vice versa without changing the result
  $$
      a \otimes (b \oplus c) = (a \otimes b) \oplus (a \otimes c)   ~ \Rightarrow ~
      \mathbf{A} \otimes (\mathbf{B} \oplus \mathbf{C}) = (\mathbf{A} \otimes \mathbf{B}) \oplus (\mathbf{A} \otimes \mathbf{C})
  $$

Matrix multiply distributivity allows graphs to be transformed via matrix multiply and then combined or vice versa without changing the result
  $$
      a \otimes (b \oplus c) = (a \otimes b) \oplus (a \otimes c)   ~~~ \Rightarrow ~~~
      \mathbf{A} (\mathbf{B} \oplus \mathbf{C}) = (\mathbf{A} \mathbf{B}) \oplus (\mathbf{A} \mathbf{C})
  $$
 where matrix multiply
 $$
   \mathbf{C} = \mathbf{A} {\oplus}.{\otimes} \mathbf{B} = \mathbf{A} \mathbf{B}
$$
is given by
  $$
   {\bf C}(i,j) = \bigoplus_{k=1}^l {\bf A}(i,k) \otimes {\bf B}(k,j)
  $$
for matrices with dimensions
$$
  {\bf A} : \mathbb{S}^{m \times l} ~~~~~
  {\bf B} : \mathbb{S}^{l \times m} ~~~~~
  {\bf C} : \mathbb{S}^{m \times n}
$$

Matrix multiply associativity is another implication of scalar distributivity and allows graphs to be transformed via matrix multiplication in various orderings without changing the result
  $$
      a \otimes (b \oplus c) = (a \otimes b) \oplus (a \otimes c)   ~~~~~ \Rightarrow ~~~~~
      (\mathbf{A} \mathbf{B}) \mathbf{C} = \mathbf{A} (\mathbf{B} \mathbf{C})
  $$

Matrix multiply commutativity can be achieved when combined with the transpose operation
$$
  (\mathbf{A} \mathbf{B})^{\sf T} = \mathbf{B}^{\sf T} \mathbf{A}^{\sf T}
$$
where the transpose of a matrix is given by
$$
  \mathbf{A}^{\sf T}(j,i) = \mathbf{A}(i,j)
$$


\section{0-Element: No Graph Edge}
Sparse matrices play an important role in graphs.  Many implementations of sparse matrices reduce storage by not storing the 0-valued elements in the matrix.  In adjacency matrices, the 0 element is equivalent to no edge from the vertex that is represented by the row to the vertex that is represented by the column. In incidence matrices, the 0 element is equivalent to the edge represented by the row not including the vertex that is represented by the column.  In most cases, the 0 element is standard arithmetic 0, but in other cases it can be a different value.  Nonstandard 0 values can be helpful when combined with different $\oplus$ and $\otimes$ operations.  For example, in different contexts 0 might be $+\infty$, -$\infty$, or $\emptyset$ (empty set).
For any value of 0, if the 0 element has certain properties with respect to scalar $\oplus$ and $\otimes$, then the sparsity of matrix operations can be managed efficiently.  These properties are the additive identity
$$
     a \oplus 0 = a
$$
and the multiplicative annihilator
$$
     a \otimes 0 = 0
$$

  Example combinations of $\oplus$ and $\otimes$ that exhibit the additive identity and multiplicative annihilator include
\begin{itemize}
\item standard arithmetic  (${+}.{\times}$) on real numbers $\mathbb{R}$
\item max-plus algebra (${\max}.{+}$) on real numbers with a defined minimal element $\{\text{-}\infty \cup \mathbb{R}\}$
\item min-plus algebra (${\min}.{+}$) using real numbers with a defined maximal element $\{\mathbb{R} \cup \infty\}$
\item max-min algebra (${\max}.{\min}$) using non-negative real numbers $[0,\infty)$
\item min-max algebra (${\min}.{\max}$)] using non-positive real numbers $(\text{-}\infty,\leq 0]$
\item max-min algebra (${\max}.{\min}$) using non-positive real numbers with a minimal element $\{\text{-}\infty \cup \mathbb{R}_{\leq 0} \}$
\item min-max algebra (${\min}.{\max}$) using non-negative real numbers with a maximal element $\{\mathbb{R}_{\geq 0} \cup \infty\}$
\item Galois field (${{\rm xor}}.{{\rm and}}$) over a set of two numbers $\{0,1\}$
\item power set (${\cup}.{\cap}$)] on any subset of integers $\mathbb{Z}$
\end{itemize}

The above examples are a small selection of the operators and sets that are useful for building graph algorithms.  Many more are possible.  The ability to change the  scalar values and operators while preserving the overall behavior of the graph operations is one of the principal benefits of using matrices for graph algorithms.

\section{Matrix Graph Operations}

The main benefit of a matrix approach to graphs is the ability to perform a wide range of graph operations on diverse types of graphs with a small number of matrix operations.  These core matrix operations and some example graph operations they support are as follows
\begin{itemize}
\item building a sparse matrix from row, column, and value triples, which corresponds to constructing a graph from a set of out-vertices, in-vertices, and edge weights
\item extracting the row, column, and value tuples corresponding to the nonzero elements in a sparse matrix, which corresponds to extracting graph edges from the matrix representation of a graph
\item transposing the rows and the columns of a sparse matrix, which is equivalent to swapping the out-vertices and the in-vertices of a graph
\item using matrix multiplication to perform single-source breadth-first search, multisource breadth-first search, and weighted breadth-first search on a graph
\item extracting a sub-matrix from a larger matrix is equivalent to selecting a sub-graph from a larger graph
\item assigning a matrix to a set of indices in a larger matrix inserts a sub-graph into a graph
\item using element-wise addition of matrices and element-wise multiplication of matrices to perform graph union and intersection along with edge weight scaling and combining
\end{itemize}
The above collection of functions has been shown to be useful for implementing a wide range of graph algorithms.  These functions strike a balance between providing enough functions to be useful to application builders while being few enough that they can be implemented effectively.

\subsection{Building a Matrix: Edge List to Graph}
Graph data can often be represented as triples of vectors ${\bf i}$, ${\bf j}$, and ${\bf v}$ corresponding to the nonzero elements in the sparse matrix.  Constructing an $m \times n$ sparse matrix from vector triples can be denoted
$$
   \mathbf{C}  ~~ {=} ~~  \mathbb{S}^{m \times n}({\bf i},{\bf j},{\bf v},\oplus)
$$
where
$$
  {\bf i} : I^l ~~~~~
  {\bf j} : J^l ~~~~~
  {\bf v} : \mathbb{S}^l
$$
are all $l$ element vectors.  The optional $\oplus$ operation defines how multiple entries with the same row and column are handled.

\subsection{Extracting Tuples: Graph to Vertex List}
Extracting the nonzero tuples from a sparse matrix can be denoted mathematically as
$$
	({\bf i},{\bf j},{\bf v}) = {\bf A}
$$

\subsection{Transpose: Swap Out-Vertices and In-Vertices}
\begin{figure}[!htb]
  \centering
    \includegraphics[width=3in]{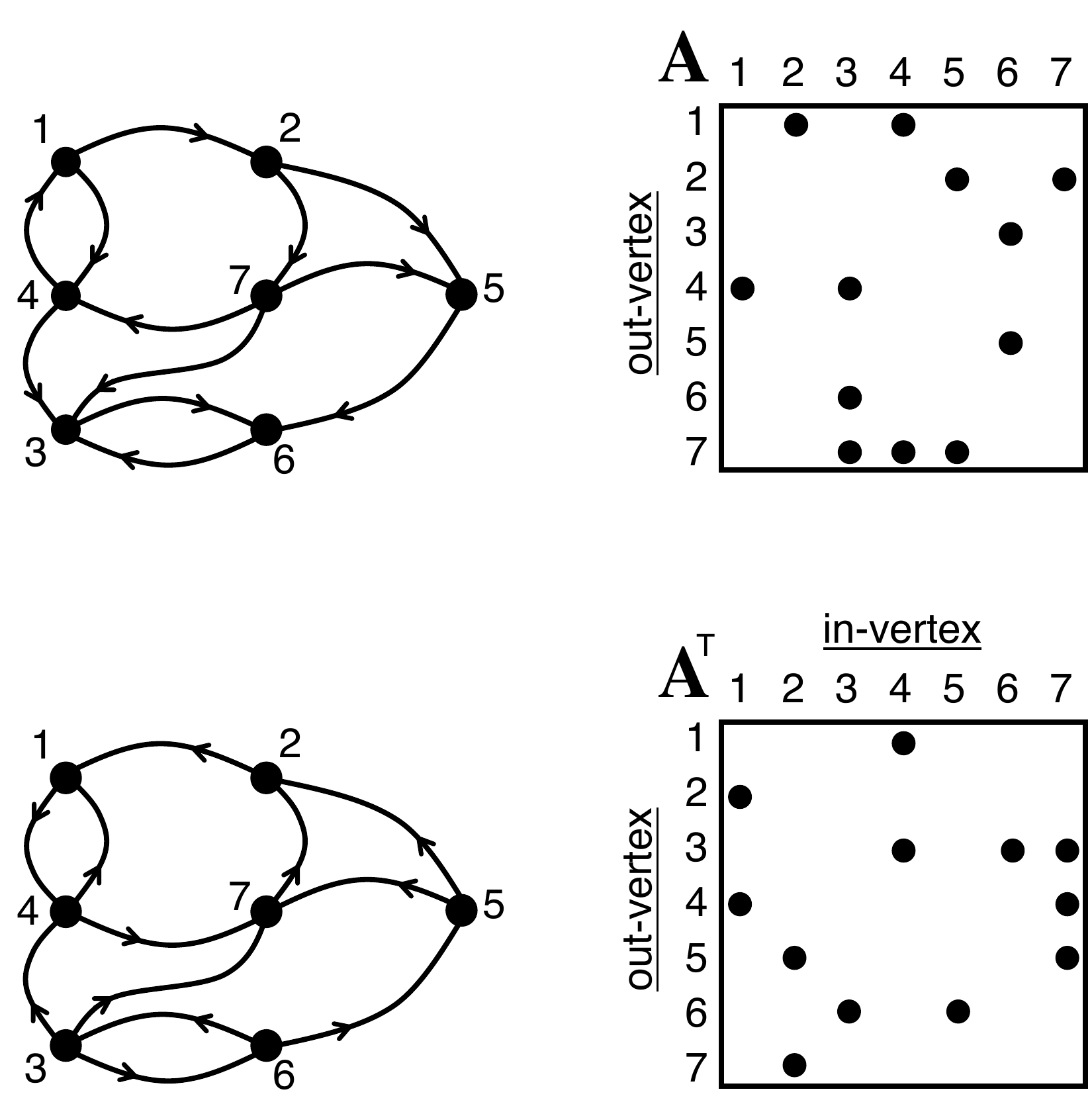}
      \caption{Transposing the adjacency matrix of a graph switches the directions of its edges.}
      \label{fig:AdjacencyMatrixTranspose}
\end{figure}

Swapping the rows and columns of a sparse matrix is a common tool for changing the direction of vertices in a graph (see Figure~\ref{fig:AdjacencyMatrixTranspose}).  The transpose is denoted as
$$
     {\bf C} ~~ {=} ~~ {\bf A}^{\sf T}
$$   
or more explicitly
$$
     {\bf C}(j,i) ~~ = ~~ {\bf A}(i,j)
$$   
where ${\bf A}:\mathbb{S}^{m \times n}$ and ${\bf C}:\mathbb{S}^{n \times m}$

Transpose also can be implemented using triples as follows
$$
	({\bf i},{\bf j},{\bf v}) = {\bf A}
$$
$$
   \mathbf{C} = \mathbb{S}^{n \times m}({\bf j},{\bf i},{\bf v})
$$

\subsection{Matrix Multiplication: Breadth-First-Search, and Adjacency Matrix Construction}
\begin{figure}[!htb]
  \centering
    \includegraphics[width=3in]{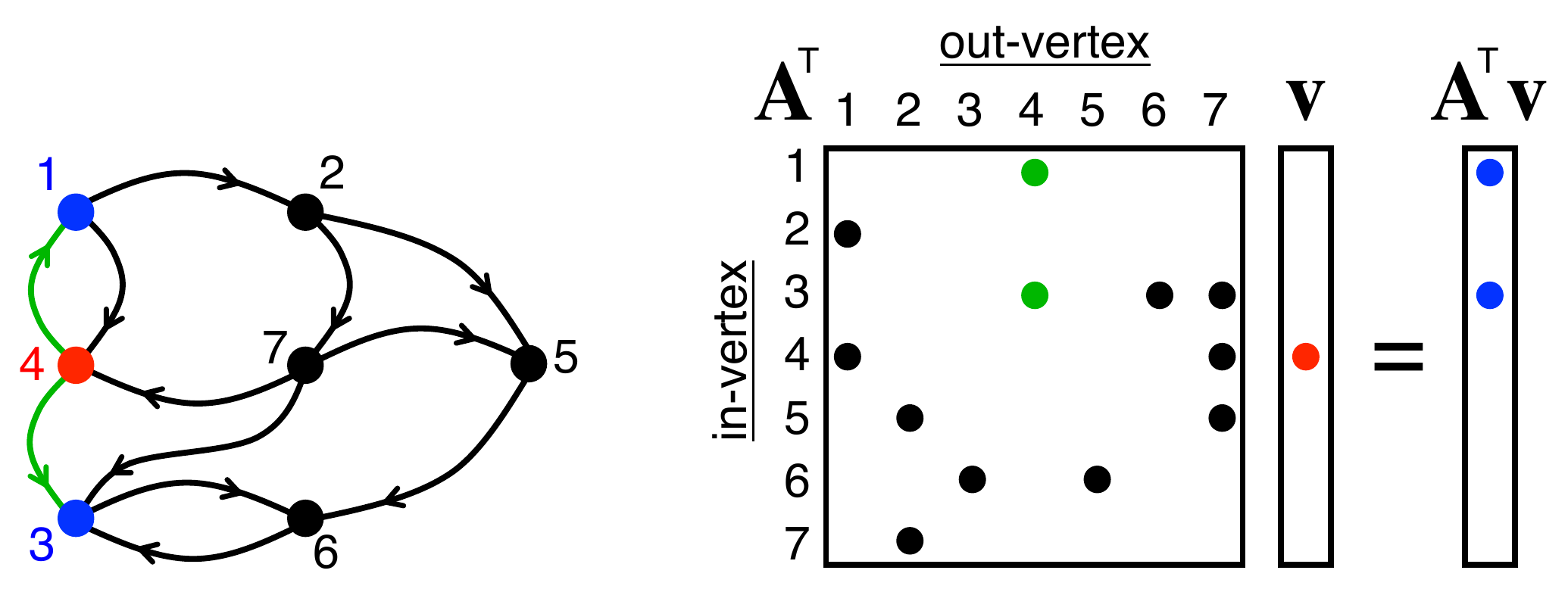}
      \caption{(left) Breadth-first search of a graph starting at vertex 4 and traversing to vertices 1 and 3.  (right) Matrix-vector multiplication of the adjacency matrix of a graph performs the equivalent operation.}
      \label{fig:AdjacencyMatrixBFS}
\end{figure}

\begin{figure}[!htb]
  \centering
    \includegraphics[width=3in]{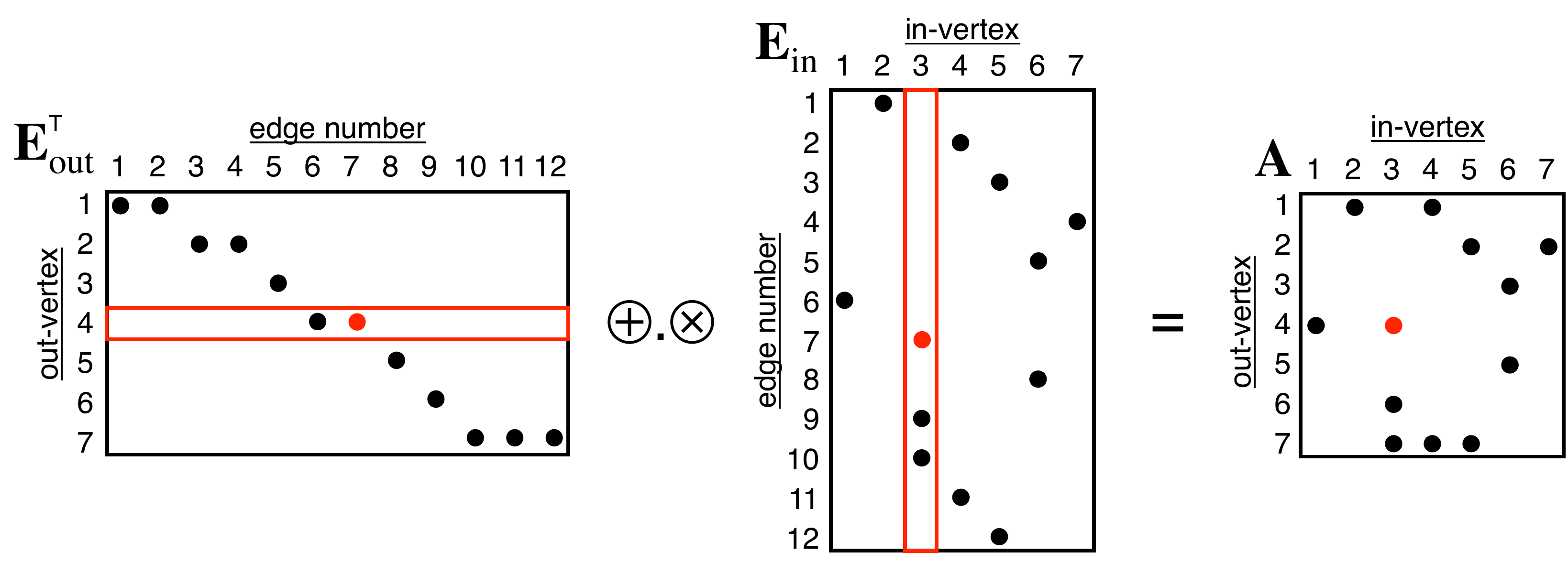}
      \caption{Construction of an adjacency matrix of a graph from its incidence matrices via matrix-matrix multiply.  The entry $\mathbf{A}(4,3)$ is obtained by combining the row vector $\mathbf{E}_\mathrm{out}^{\sf T}(4,k)$ with the column vector $\mathbf{E}_\mathrm{in}(k,3)$ via matrix-matrix product 
$
  \mathbf{A}(4,3)= \bigoplus\limits_{k = 1}^{12} \mathbf{E}_\mathrm{out}^{\sf T}(4,k) \otimes \mathbf{E}_\mathrm{in}(k,3)
$.}
      \label{fig:AdjacencyToIncidence}
\end{figure}

Matrix multiplication is the most important matrix operation and can be used to implement a wide range of graph algorithms.  Examples include finding the nearest neighbors of a vertex (see Figure~\ref{fig:AdjacencyMatrixBFS}) and constructing an adjacency matrix from an incidence matrix (see Figure~\ref{fig:AdjacencyToIncidence}).  In its most common form, matrix multiplication using standard arithmetic addition and multiplication is given by
$$
   {\bf C} = {\bf A} {\bf B}
$$
or more explicitly
$$
   {\bf C}(i,j) = \sum_{k=1}^l {\bf A}(i,k) {\bf B}(k,j)
$$
where
$$
  {\bf A} : \mathbb{R}^{m \times l} ~~~~~
  {\bf B} : \mathbb{R}^{l \times n} ~~~~~
  {\bf C} : \mathbb{R}^{m \times n}
$$
Matrix multiplication has many important variants that include non-arithmetic addition and multiplication
$$
   {\bf C} ~~ {=} ~~ {\bf A} ~ {\oplus}.{\otimes} ~ {\bf B}
$$
where
$$
  {\bf A} : \mathbb{S}^{m \times l} ~~~~~
  {\bf B} : \mathbb{S}^{l \times n} ~~~~~
  {\bf C} : \mathbb{S}^{m \times n}
$$
and the notation ${\oplus}.{\otimes}$ makes explicit that ${\oplus}$ and ${\otimes}$ can be other functions.

One of the most common uses of matrix multiplication is to construct an adjacency matrix from an incidence matrix representation of a graph.  For a graph with out-vertex incidence matrix $\mathbf{E}_\mathrm{out}$ and in-vertex incidence matrix $\mathbf{E}_\mathrm{in}$, the corresponding adjacency matrix can be computed by
$$
  \mathbf{A} = \mathbf{E}_\mathrm{out}^{\sf T} \mathbf{E}_\mathrm{in}
$$
where the individual values in $\mathbf{A}$ can be computed via
$$
  \mathbf{A}(i,j)= \bigoplus\limits_{k} \mathbf{E}_\mathrm{out}^{\sf T}(i,k) \otimes \mathbf{E}_\mathrm{in}(k,j)
$$


\subsection{Extract: Selecting Sub-graphs}
\begin{figure}[!htb]
  \centering
    \includegraphics[width=3in]{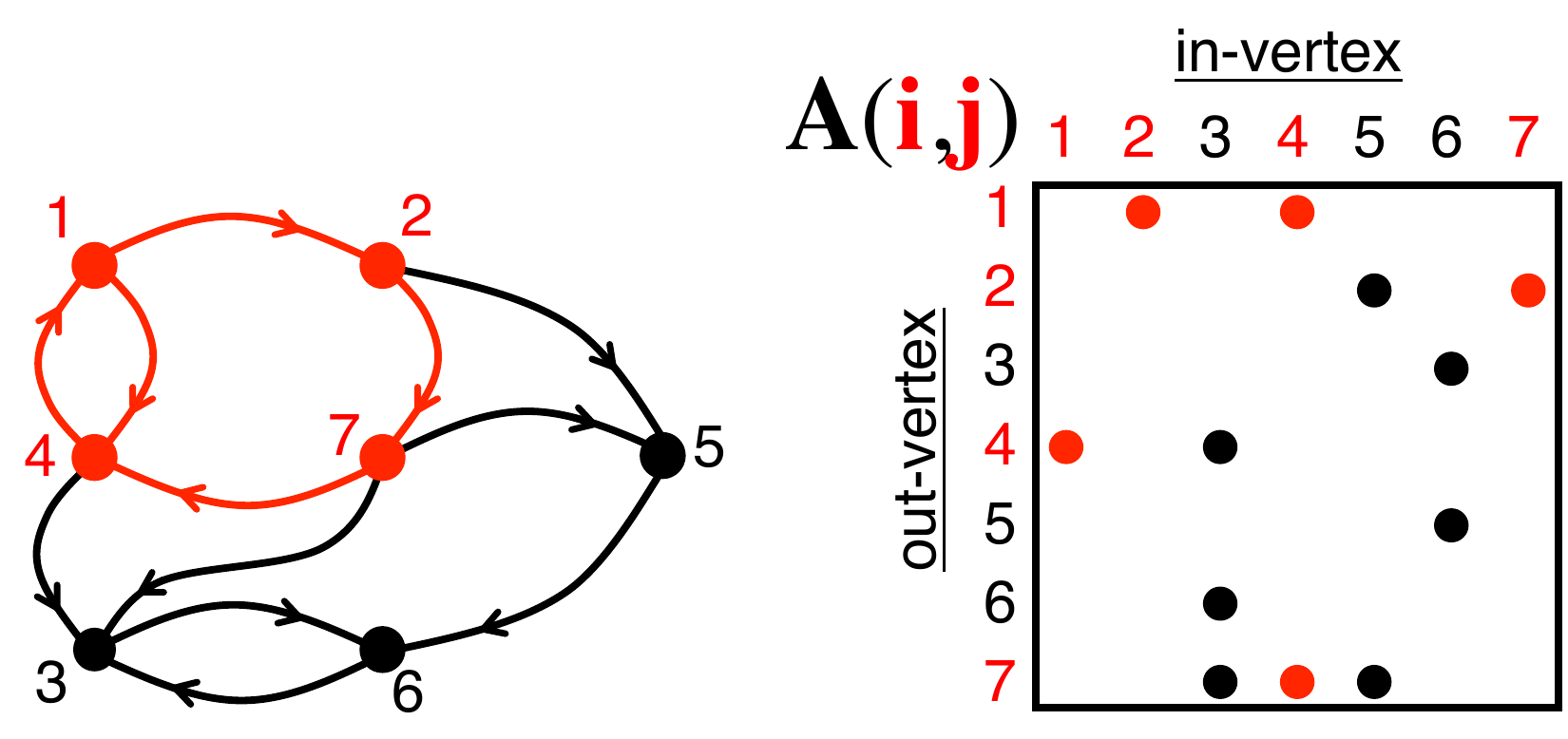}
      \caption{Selection of a 4-vertex sub-graph from the adjacency matrix via selecting subsets of rows and columns ${\bf i} = {\bf j} = \{1,2,4,7\}$.}
      \label{fig:AdjacencyMatrixSub}
\end{figure}

  Selecting sub-graphs is a very common graph operation (see Figure~\ref{fig:AdjacencyMatrixSub}).  This operation is performed by selecting out-vertices (row) and in-vertices (columns) from a matrix $\mathbf{A} : \mathbb{S}^{m \times n}$
$$
   \mathbf{C} ~~ {=} ~~  \mathbf{A}({\bf i},{\bf j})
$$
or more explicitly
$$
   {\bf C}(i,j) = {\bf A}({\bf i}(i),{\bf j}(j))
$$
where $i \in \{1,...,m_C\}$, $j \in \{1,...,n_C\}$, ${\bf i} : I^{m_C}$, and ${\bf j}: J^{m_C}$  select specific sets of rows and columns in a specific order.   The resulting matrix $\mathbf{C} : \mathbb{S}^{m_C \times n_C}$ can be larger or smaller than the input matrix $\mathbf{A}$.  This operation can also be used to replicate and/or permute rows and columns in a matrix.

  Extraction can also be implemented with matrix multiplication as
$$
   \mathbf{C}= \mathbf{S}({\bf i}) ~ \mathbf{A} ~ \mathbf{S}^{\sf T}({\bf j})
$$
where $\mathbf{S}({\bf i})$ and $\mathbf{S}({\bf j})$ are selection matrices given by
$$
   \mathbf{S}({\bf i}) = \mathbb{S}^{m_C \times m}(\{1,...,m_C\},{\bf i},1)
$$
$$
    \mathbf{S}({\bf j}) = \mathbb{S}^{n_C \times n}(\{1,...,n_C\},{\bf j},1)
$$

\subsection{Assign: Modifying Sub-Graphs}
  Modifying sub-graphs is a very common graph operation.  This operation is performed by selecting out-vertices (row) and in-vertices (columns) from a matrix $\mathbf{C} : \mathbb{S}^{m \times n}$ and assigning new values to them from another sparse matrix, $\mathbf{A} : \mathbb{S}^{m_A \times n_A}$
$$
   \mathbf{C}({\bf i},{\bf j}) = \mathbf{A}
$$
or more explicitly
$$
   {\bf C}({\bf i}(i),{\bf j}(j)) = {\bf A}(i,j) 
$$
where $i \in \{1,...,m_A\}$, $j \in \{1,...,n_A\}$, ${\bf i} : I^{m_A}$ and ${\bf j}: J^{n_A}$ select specific sets of rows and columns. 


\subsection{Element-Wise Addition and Element-Wise Multiplication: Combining Graphs, Intersecting Graphs, and Scaling Graphs}
  Combining graphs along with adding their edge weights can be accomplished by adding together their sparse matrix representations
$$
   {\bf C} = {\bf A} \oplus {\bf B}
$$
where ${\bf A},{\bf B},{\bf C}: \mathbb{S}^{m \times n}$ or more explicitly 
$$
   {\bf C}(i,j) = {\bf A}(i,j)  \oplus {\bf B}(i,j)
$$
where $i \in \{1,...,m\}$, and $j \in \{1,...,n\}$. 

  Intersecting graphs along with scaling their edge weights can be accomplished by element-wise multiplication of their sparse matrix representations
$$
   {\bf C} = {\bf A} \otimes {\bf B}
$$
where ${\bf A},{\bf B},{\bf C}: \mathbb{S}^{m \times n}$ or more explicitly 
$$
   {\bf C}(i,j) = {\bf A}(i,j) \otimes{\bf B}(i,j)
$$
where $i \in \{1,...,m\}$, and $j \in \{1,...,n\}$.

\section{Performance}


A standard such as the GraphBLAS can only be effective if it does not
impose unnecessary overhead on the computations it performs. One test
of the overhead is to compare the GraphBLAS implementation to other
standard sparse matrix libraries. Figure 10 shows the performance of
one prototype GraphBLAS implementation compared to a state-of-the art
GPU graph library (Gunrock)~\cite{Wang2016}.

The dataset used are random undirected Kronecker graphs with edge
factor 32 and scale factor ranging from 16 to 21. Each experiment
conducts a BFS starting from a high degree node in the graph. The
GraphBLAS performance of sparse matrix - sparse vector multiplication 
is similar to Gunrock BFS performance. The similarity in
performance indicates that the GraphBLAS is not introducing a high
overhead. Each experiment is launched on these graphs from node 0
except on the scale 19 graph, which is launched from node 1. The
runtime is an average of 10 runs to reduce variance.

We ran all experiments in this paper on a Linux workstation with $2\times$
3.50 GHz Intel 4-core E5-2637 v2 Xeon CPUs, 256 GB of main memory, and
an NVIDIA K40c GPU with 12 GB on-board memory. The GPU programs were
compiled with NVIDIA's nvcc compiler (version 7.5.17) using the \texttt{-O3}
optimization level. The C code was compiled using gcc 4.8.5. All
results ignore transfer time (from disk-to-memory and CPU-to-GPU). The
Gunrock code was executed using the command-line configuration
\texttt{--undirected --traversal-mode=1 --iteration-num=10}.

\begin{figure}[!htb]
  \centering
    \includegraphics[width=3in]{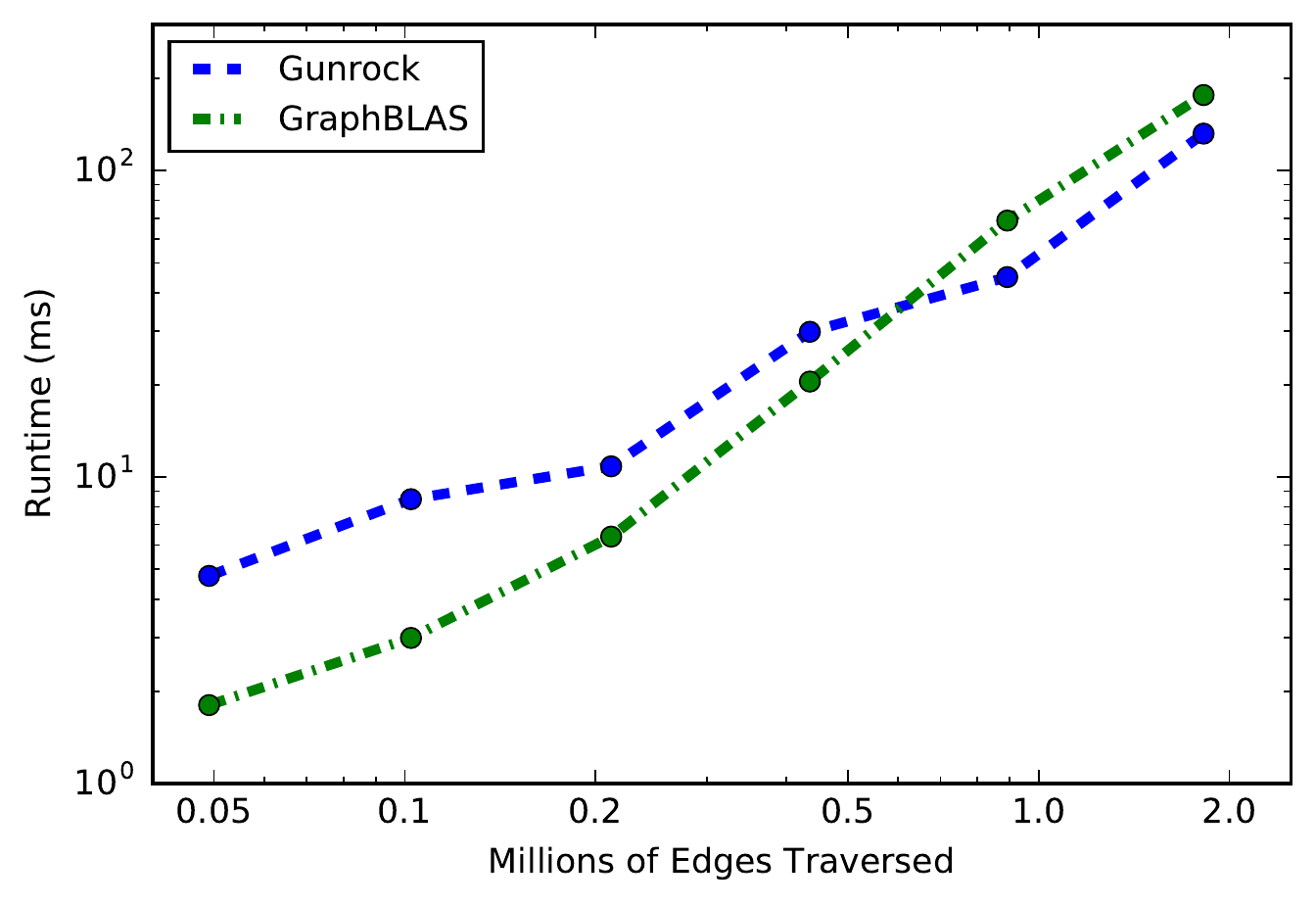}
      \caption{Sparse matrix times sparse vector performance for a prototype GraphBLAS implementation as compared to an optimized GPU graph library (Gunrock) performing BFS in a similar manner.}
      \label{fig:Performance}
\end{figure}

\begin{figure}[!htb]
  \centering
    \includegraphics[width=3in]{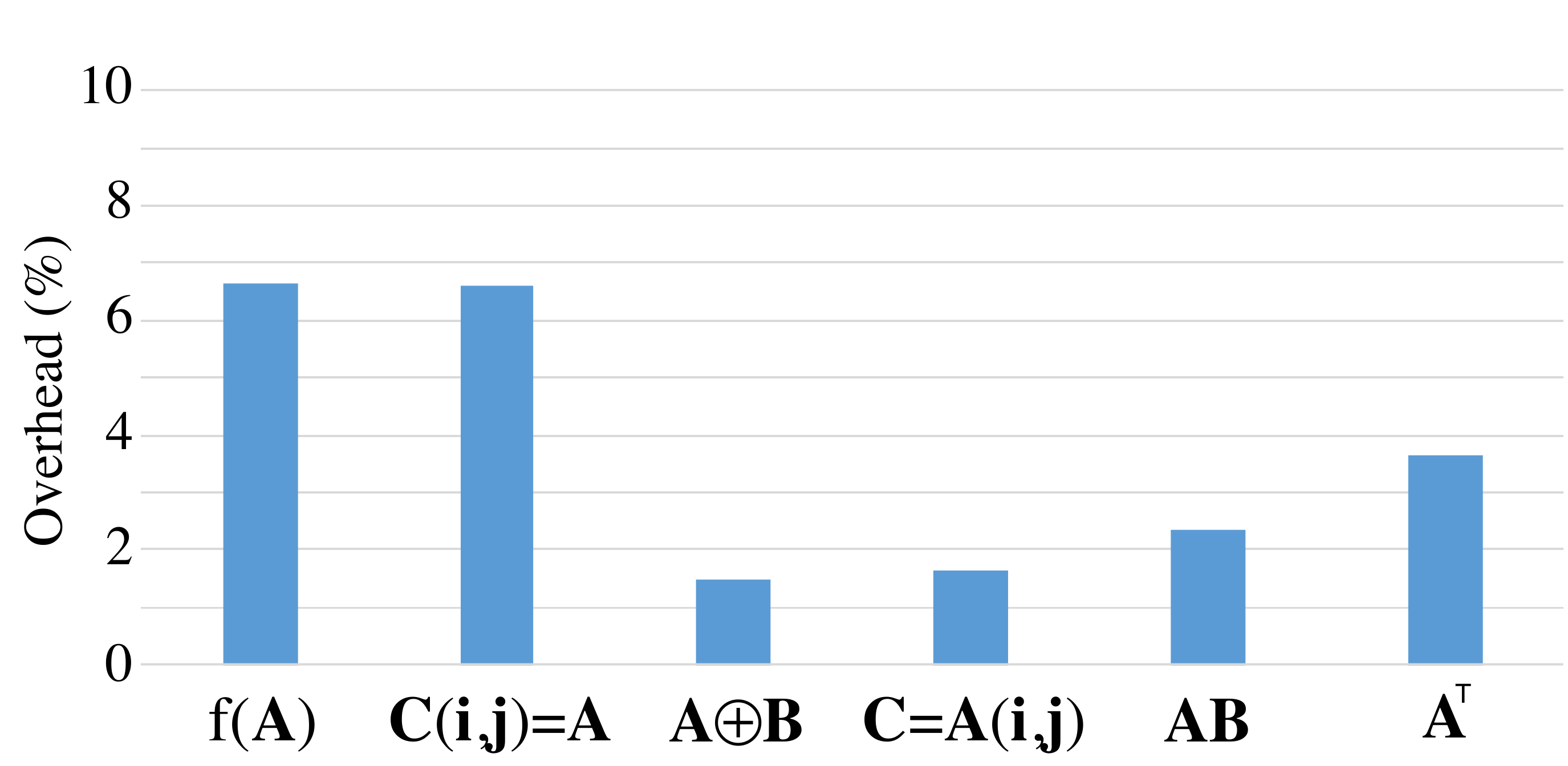}
      \caption{Percentage overhead of the GraphBLAS Template Library prototype implementation on six different GraphBLAS operations.}
      \label{fig:OverheadPercentage}
\end{figure}

  Figure~\ref{fig:OverheadPercentage} shows the overhead of a second prototype GraphBLAS implementation, the GraphBLAS Template Library (GBTL)\cite{Zhang2016}.
  We measured the GraphBLAS API overhead using the GraphBLAS Template Library (GBTL) on a machine with an Intel i5-4670k processor and a GTX660 CUDA-capable graphics card. The overhead results reflect the difference in runtime, in terms of percentages, between the CUDA backend of GBTL invoked using GraphBLAS API and the direct calling of underlying implementation. We obtain the numbers by averaging the overhead of 16 runs on Erd\H os-R\'{e}nyi random graphs generated using the same dimension and sparsity. The code is compiled using the \texttt{-O2} optimization level on version 7.5.18 of the CUDA toolkit with gcc 4.9.3. 
  The results indicate that the overhead of the GraphBLAS is small compared to the underlying math being performed.



\section{Conclusions}

Matrices are a powerful tool for representing and manipulating graphs.  Adjacency matrices represent directed-weighted-graphs with each row and column in the matrix representing a vertex and the values representing the weights of the edges.  Incidence matrices represent directed-weighted-multi-hyper-graphs with each row representing an edge and each column representing a vertex.  Perhaps the most important aspects of matrix-based graphs are the mathematical properties of commutativity, associativity, and distributivity.  These properties allow a very small number of matrix operations to be used to construct a large number of graphs.  These properties of the matrix are determined by the element-wise properties of addition and multiplication on the values in the matrix. The GraphBLAS allows these matrix properties to be readily applied to graphs in a low-overhead manner.

\section*{Acknowledgments}

The authors would like to thank Hedayat Alghassi, Michael Anderson, Ariful Azad, Muthu Baskaran, Paul Burkhardt, Steven Dalton, Tim Davis, Joe Eaton, Alan Edelman, Sterling Foster, Vijay Gadepally, Joseph Gonzalez, Torsten Hoefler, Erik Holk, Thejaka Kanewala, Tze Meng Low,  Dave Martinez, John Matty, Asit Mishra, Samantha Misurda, Mostofa Patwary, Fabrizio Petrini, Albert Reuther, Jason Riedy, Victor Roytburd, Nadathur Satish, Narayanan Sundaram, Richard Veras, Michael Wolf, Albert-Jan Yzelman,  Peter Zhang, and Xia Zhu.



%
%
%

\end{document}